\documentclass[pre,twocolumn,nofootinbib,showpacs]{revtex4-1}
\pdfoutput=1
\usepackage[english]{babel}
\usepackage[utf8]{inputenc}
\usepackage{graphicx}
\usepackage{hyperref}
\usepackage{gensymb}
\usepackage{setspace}
\usepackage{amsmath}

\begin{document}
\title{Slowdown of the surface diffusion during early stages of bacterial colonization}

\author{T. Vourc'h}
\affiliation{Laboratoire AstroParticules et Cosmologie, CNRS,  Université Paris-Diderot,  Université Sorbonne Paris Cité, 5 rue Thomas Mann 75013 Paris, France}

\author{J. L\'eopold\`es}

\email{julien.leopoldes@espci.fr }
\affiliation{ESPCI Paris, PSL Research University, CNRS, Institut Langevin, 1 rue Jussieu, F-75005 Paris, France}
\affiliation{Université Paris-Est Marne-la-Vall\'ee, 5 Bd Descartes, Champs sur Marne, Marne-la-Vall\'ee cedex 2, France}

\author{A. M\'ejean}
\affiliation{Laboratoire Interdisciplinaire des \'Energies de Demain, CNRS,  Universit\'e Paris-Diderot,  Universit\'e Sorbonne Paris Cit\'e, 5 rue Thomas Mann 75013 Paris, France}
\author{H. Peerhossaini}
\email{hassan.peerhossaini@univ-paris-diderot.fr }
\affiliation{Laboratoire AstroParticules et Cosmologie, CNRS,  Université Paris-Diderot,  Université Sorbonne Paris Cité, 5 rue Thomas Mann 75013 Paris, France}

\author{F. Chauvat}
\affiliation{Laboratory of Biology and Biotechnology of Cyanobacteria. Institute for Integrative Biology of the Cell (I2BC), CEA, CNRS, Univ. Paris-Sud, Université Paris-Saclay, 91198, Gif-sur-Yvette cedex, France.
 }
\author{C. Cassier-Chauvat}
\affiliation{Laboratory of Biology and Biotechnology of Cyanobacteria. Institute for Integrative Biology of the Cell (I2BC), CEA, CNRS, Univ. Paris-Sud, Université Paris-Saclay, 91198, Gif-sur-Yvette cedex, France.
 }


\date{\today}

\begin{abstract}
We study the surface diffusion of the model cyanobacterium \textit{Synechocystis} sp. PCC $6803$ during the incipient stages of cell contact with a glass surface in the dilute regime. We observe a twitching motility with alternating immobile ``tumble" and mobile ``run" periods, resulting in a normal diffusion described by a continuous time random walk with a coefficient of diffusion $D$. Surprisingly, $D$ is found to decrease with time down to a plateau. This is observed only when the cyanobacterial cells are able to produce released extracellular polysaccharides, as shown by a comparative study between the wild-type strain and various polysaccharides-depleted mutants. The analysis of the trajectories taken by the bacterial cells shows that the temporal characteristics of their intermittent motion depend on the instantaneous fraction of visited sites during diffusion. This describes quantitatively the time dependence of $D$, related to the progressive surface coverage by the polysaccharides. The observed slowdown of the surface diffusion may constitute a basic precursor mechanism for microcolony formation and provides clues for controlling biofilm formation.  

\end{abstract}

\maketitle
\section{Introduction}
In nature, bacteria develop preferentially in contact with solid surfaces by forming biofilms, i.e. masses of adherent cells embedded in slimy extracellular matrices. Biofilms are essential to bacterial growth and survival to environmental stresses. They capture nutrients in the vicinity of the cells, and the peripheral cells exposed to the noxious agents protect the internal cells~\cite{Singh2017,Tuson2013}. Biofilms also develop in many industrial and medical situations and their formation is a key mechanism in the infection of a living host by pathogenic organisms \cite{Davies2003,Costerton1999,Omar2017}. 

\par The biofilm structure depends critically on mass transport, surface chemistry and surface topology \cite{Mazza2016}. The initial contact of the bacteria with the surface is followed by the formation of micro-colonies \cite{Taktikos2015}. Then, the three dimensional morphology of the mature film develops, and chemical signaling triggers the release of bacteria in the liquid medium which are then transported to other colonization sites via the flow of the liquid medium \cite{Vlamakis2013}.

\par Bacteria are known to produce high molecular weight polymeric substances such as extracellular polysaccharides (EPS)~\cite{Marshall1971-1}, that play important roles during the main stages of biofilm formation. For example, the mature biofilm contains the macromolecules adsorbed on the solid substrate which provide mechanical stability and adhesion. A distinction can be made between capsular EPS (firmly bound to the outer cell membrane) and released EPS (easily detached from the outer cell membrane) \cite{Jittawuttipoka2013}. Moreover, it has been proved recently that the polysaccharides excreted by motile bacteria form attracting trails, giving rise to spatial accumulation of the cells thereby yielding the localized growth of micro-colonies~\cite{Bhaya2013,Zhao2013,Gelimson2016}. The production of EPS is also linked to the emergence of peculiar dynamics during the initial stages of surface colonization, by altering the distribution of the velocity of bacterial cells~\cite{Hu2016}. 

In this work, we study the relationship between excreted EPS and the diffusion coefficient $D$ at the early stages of surface colonization. 
\begin{figure}[htbp]
\centering
\includegraphics[scale=0.3]{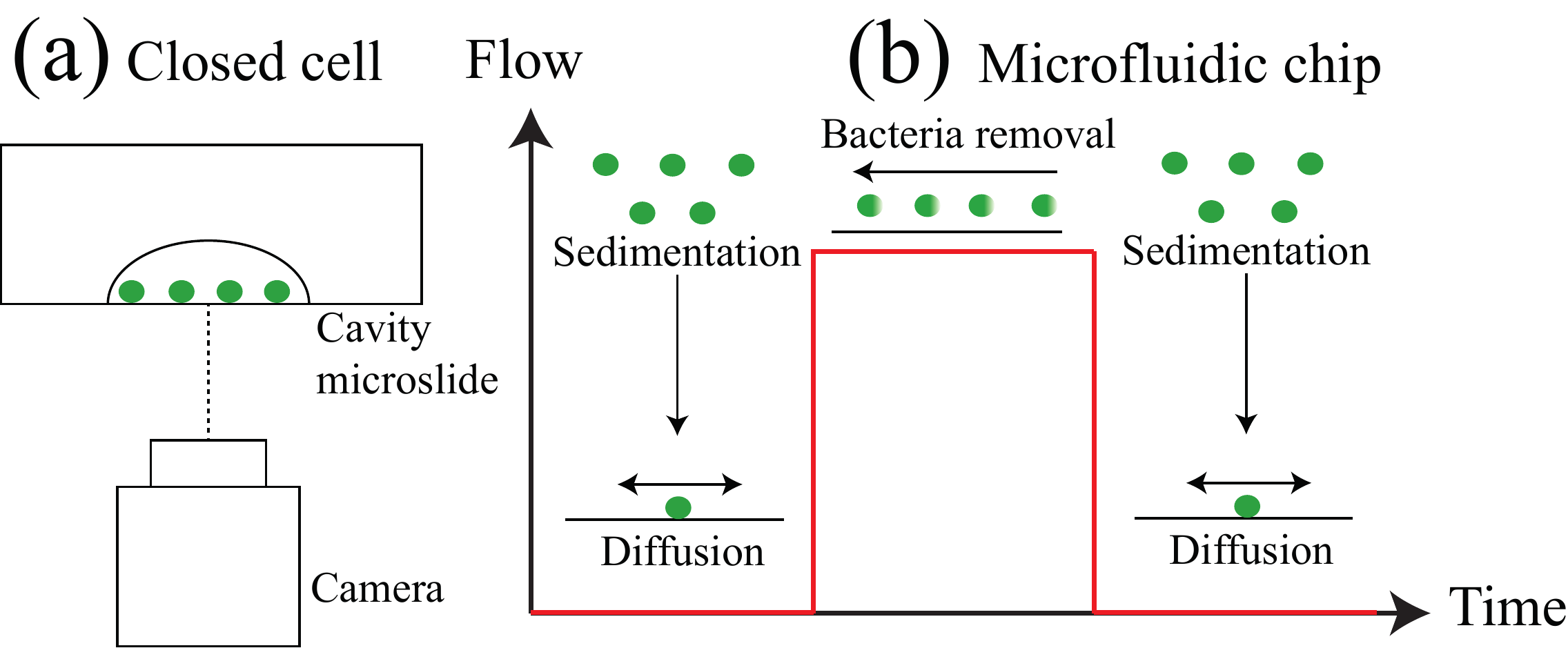}
\caption{\label{Fig1} Experimental setup (a) Closed cavity where the bacterial cells (dots) sediment and diffuse on the lower surface. (b) Protocol followed with the microfluidic cell: a first experiment is carried out similar to that in (a) but in a microfluidic cell with controlable flow. After sedimentation is completed and having let the bacteria diffuse enough on the lower surface, a  pressure gradient is applied to detach the cells from the surface. Then the flow is stopped and the sedimentation-diffusion process starts again. In both setups, bacterial cells are observed with the same equipment (optical microscope coupled to a CCD camera).}
\end{figure}
Investigations are carried out with the unicellular cyanobacterium \textit{Synechocystis} sp. PCC~$6803$, a model of environmentally important photosynthetic prokaryote that produces EPS in various forms ~\cite{Panoff1988,Jittawuttipoka2013}. The motility of \textit{Synechocystis} on solids relies on the action of type IV pili~\cite{Bhaya2000,Bhaya2002,Bhaya2013,Bhaya2015,Wilde2015} (the pili extend, bind on the solid surface and then retract). Results collected for the wild type cells are compared with those obtained with different mutant strains described elsewhere~\cite{Jittawuttipoka2013}.

\par Surface motion occurs by the usual twitching motility but the diffusion coefficient is observed to decrease systematically with time down to a plateau. This effect is observed only with the wild type and the \textit{$\Delta$sll1581} mutant strain, both able to produce released exopolysaccharides. This is not noticed for two double-mutant strains (\textit{$\Delta$sll581-sll1875} and \textit{$\Delta$sll0923-sll5052}) that produce a lower amount of released EPS level. We propose an interpretation that takes into account the coverage of the solid surface by the trails of the excreted EPS. This affects the temporal characteristics of the intermittent twitching motility of the cells. We believe that such a process constitutes an important step in the adaptation of microorganisms to hard surfaces prior to the formation of microcolonies and biofilms.  

\section{Materials and methods}
\subsection{Bacterial suspensions and measurement of cell motion}
\par The Wild Type (WT) strain of the model cyanobacterium \textit{Synechocystis}  sp. PCC $6803$ was obtained from the Pasteur Institute, while the EPS-depleted single mutant (\textit{$\Delta$sll1581}) and double mutants (\textit{$\Delta$sll581-sll1875} and \textit{$\Delta$sll0923-sll5052}) were previously constructed by some of us \cite{Jittawuttipoka2013}. The three mutant strains produce less capsular EPS than the WT cells. The single mutant (\textit{$\Delta$sll1581}) produces similar amounts of released EPS than the WT cells; both double mutants  (\textit{$\Delta$sll1581-sll1875}) and (\textit{$\Delta$sll0923-sll5052}) form less released EPS.
\par All strains are routinely cultured in the BG11 standard mineral medium, and sub-cultivated by diluting $3$ mL of a mother culture in 47 mL of fresh BG11. The suspensions are stirred by a magnetic agitator operating at $360$ rotations per minute in a clean room at $20^{\circ}$C. They are placed under white light intensity of 1.3~W.m$^{-2}$ for 7 days followed by 24 hours dark and subsequent 2 hours light before running the experiments. At this stage, the concentration of cyanobacteria is approximately $2.10^7$ cells per mL. The suspensions are diluted 2-to10-fold in fresh BG11 before introduction in the measurement chamber. With this protocol, some of the cells are dividing, which results in average particle diameter $d\sim 3$~$\mu$m, slightly larger that single cells, whose size is approximately $2$~$\mu$m.

\par Experiments are conducted in two different systems as represented in Figure~\ref{Fig1}. One measurement cell consists in a BRAND$^{\textregistered}$ cavity microscope slide (26$\times$76~mm) closed by a cover slip (Menzel-Gläser, 22$\times$22~mm) and sealed with high vacuum grease (Figure~\ref{Fig1}(a)). The second measurement cell is a Y-junction microfluidic channel of cross section $100\times 205$~$\mu$m, see protocol detailed in Figure~\ref{Fig1}(b).

\subsection{Cell tracking}
The cyanobacterial cells are observed with a home made inverted microscope equipped with a Nikon TU Plan 10X objective and a white light source. Their motion is recorded with a monochrome camera (Edmund Optics) at one frame per second. The recordings are post-processed with ImageJ software to obtain binary images and then are analyzed for particle tracking \cite{Lien} with MATLAB. Only trajectories whose duration are longer than 250 seconds are retained for further analysis, while the few non-motile cells are ignored. The number of analyzed trajectories is $9843$ for section~\ref{Transients}, $453$ for the section~\ref{Permanent}, and $1424$ for the experiments in the microfluidic chip. Details on the computation of the mean square displacement (MSD) are given in Appendix~\ref{app_msd}.

\begin{figure}
\includegraphics[scale=0.35]{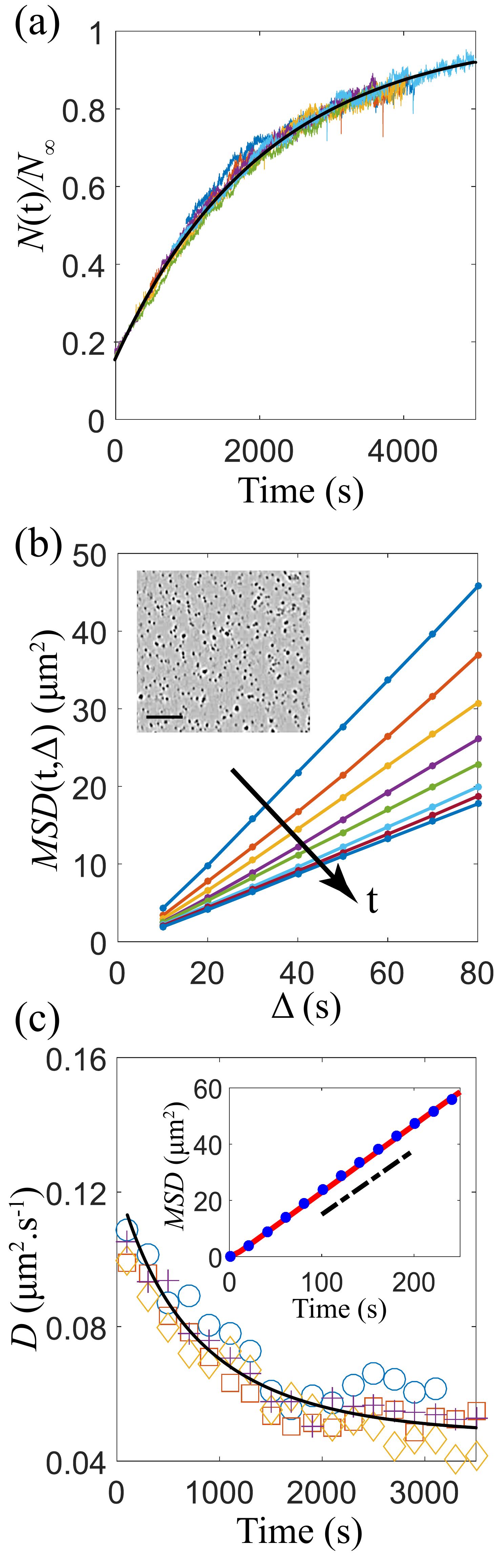}
\caption{\label{Base} (a) For five different experiments, temporal evolution of the number of cyanobacterial cells detected on the hard surface divided by the final number of cells. Plain line corresponds to Eq.~\ref{Npart}. (b) Mean square displacement (MSD) at several observation times for a selected experiment, computed according to Eq.~\ref{MSDdt}. The MSD is plotted for observation times ranging from $t=100$ to $2900$~s by step of $400$~s. Increasing  observation time is indicated by the arrow. Inset: snapshot of a typical experiment where bacteria appear as dark spots on the grey background (scale bar $50$~$\mu$m). (c) Symbols: temporal evolution of the diffusion coefficient for experiments similar to (a); Black line: fit based on Eq.~\ref{eqD} and Eq.~\ref{sdet}, see discussion section for the fitting procedure. Inset: experimental MSD at long times (circles) and as computed from numerical simulations (line). The dashed black line indicates the slope given by Eq.~\ref{eqD}.}
\end{figure}

\section{Results}

\subsection{\label{Transients}Cell transport, contact with solid surface and slowdown of motion}
\begin{figure}
\includegraphics[width=9 cm]{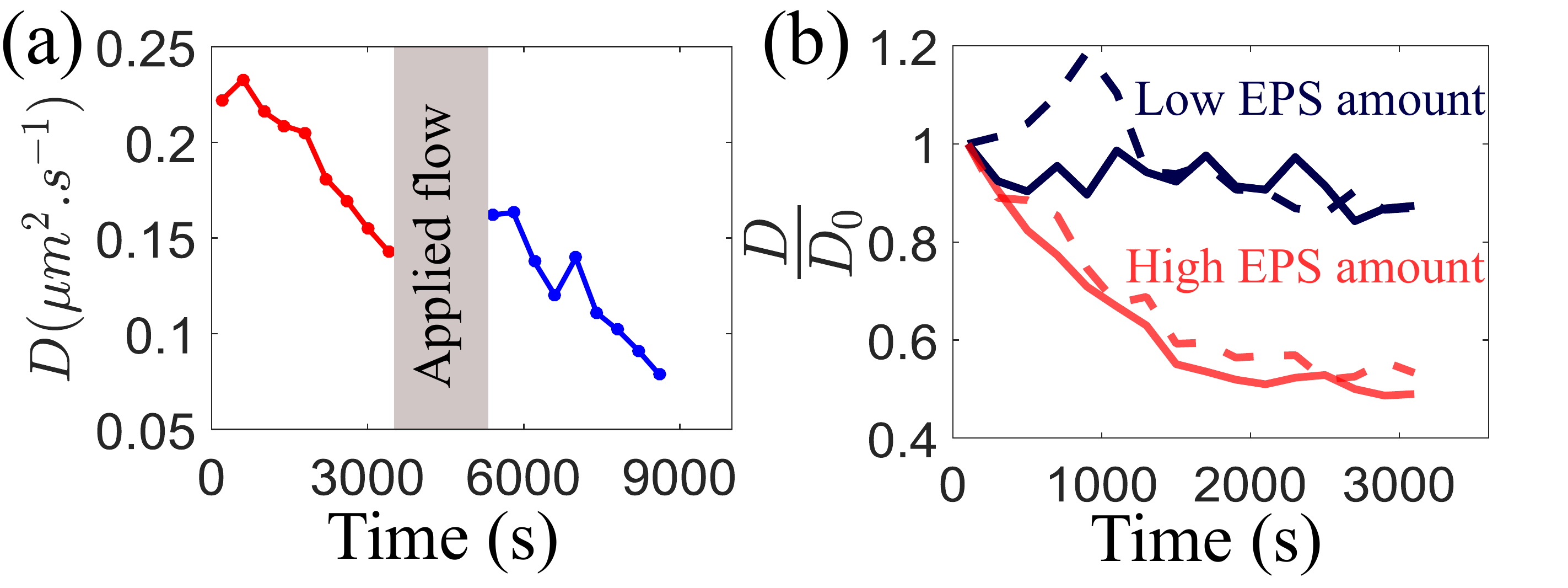}
\caption{\label{Fig6} (a) Temporal evolution of the coefficient of diffusion for experiments carried out in the microfluidic system, see Figure~\ref{Fig1}(b). Renewal of the bacterial cells population due to the liquid flow occurs during the period corresponding to the vertical grey bar. (b) Temporal evolution of the diffusion coefficient normalized by its initial value $D_0$ for various \textit{Synechocystis} strains. Data are the result of averaging over two different experiments. Dark dashed line: $\Delta$s$ll1875$-sll$1581$, dark plain line: $\Delta$sll0923-sll5052, light plain line: Wild Type and light dashed line: $\Delta$sll1581.}
\end{figure}
Data recording starts a short time after the introduction of the cyanobacterial cells in the measurement chamber. Due to cell sedimentation, the number $N(t)$ of bacterial cells detected on the hard surface increases with observation time $t$ until a final value $N_{\infty}$ is reached, see Fig.\ref{Base}(a). The sedimentation process is reproducible and can be described by the empirical law 
\begin{equation}
\label{Npart}
N(t)=N_0+(N_\infty-N_0) (1-e^{-t / \tau_N}) 
\end{equation}
where $\tau_N = 2081 \pm 4$~s and $N_0$ is the initial number of bacteria at the surface. The characteristic time for sedimentation $\tau_N$ can be obtained from the Stokes velocity $v_S=\Delta \rho g V_p/6\pi \eta R$, where $\Delta \rho$ is the density contrast between the bacterial cells and the culture medium, $g$ is the acceleration of gravity, $V_p$ is the volume of the bacterial cells, $R$ is their radius and $\eta \approx 10^{-3}$ Pa.s is the dynamical viscosity of the suspension as measured by means of a horizontal capillary. With $\Delta \rho=100$ ~$kg.m^{-3}$ \cite{Glaser1989}, $g=9.81$ m$.s^{-2}$,  $R=1.5$~$\mu$m and the height of the cell $h=1$~mm, we obtain $\tau_N \sim 2000$ s, consistent with experiments.

\begin{figure*}
\includegraphics[scale=0.35]{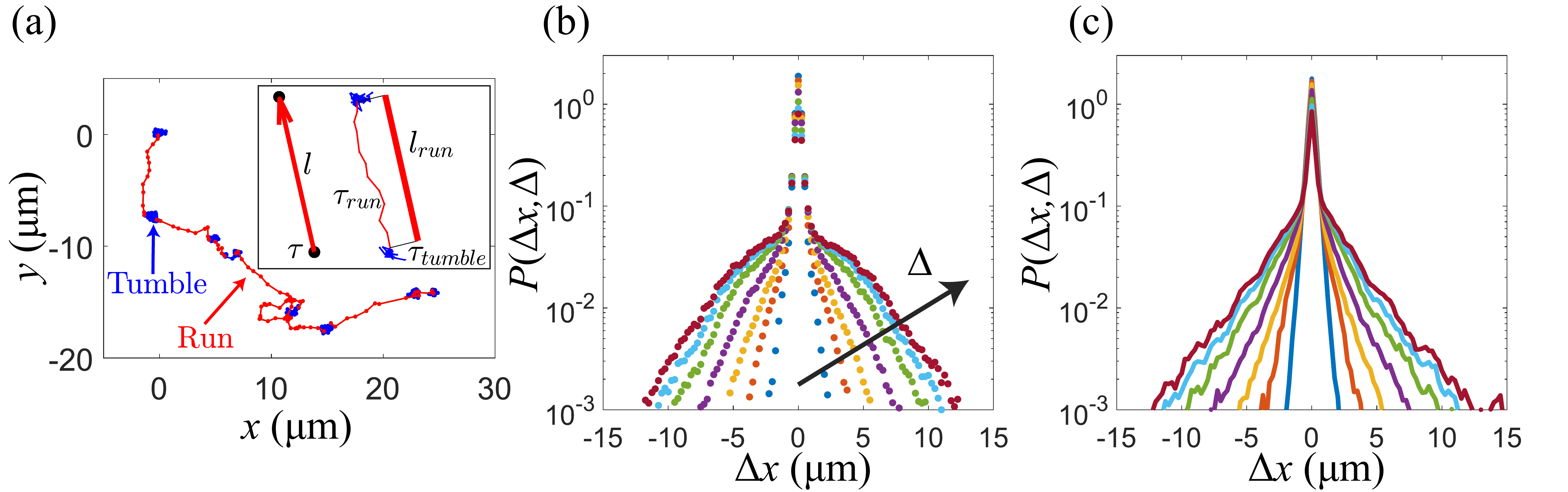}
\caption{\label{Traj2c}(a) Trajectory of a bacterial cell ($668$~s). Runs correspond to lines (red online) and tumble to spots (blue online). Inset: left, representation of a diffusion step by the continuous time random walk model and right, the corresponding experimental diffusion step where the distance travelled during a run $l_{run}$, the corresponding run duration $\tau_{run}$ and the tumble duration $\tau_{tumble}$ are indicated. (b) and (c) are displacement probabilities along one direction for experimental trajectories at the plateau and simulated trajectories, respectively. The distributions are given for time intervals $\Delta=2,5,10,20,40,60$ and $80$ s. Increasing time interval is indicated by the arrow.}
\end{figure*}

\par The dynamics of surface motion is first described by computing the mean square displacement during a short time interval $10<\Delta<80$ s, as a function of the observation time $t$ (Eq.~\ref{MSDdt}, Appendix~\ref{app_msd}). Figure~\ref{Base}(b) indicates that this short-time MSD is a linear function of $\Delta$ for all observation times $t$, as for Fickian diffusion. However, the corresponding time-dependent diffusion coefficient $D(t)$ defined by $MSD(t,\Delta)=4D(t)\Delta$ decreases with $t$. Such a slowdown of the dynamics, which constitutes the main result of this work, is reported in Fig.\ref{Base}(c) for different experiments, showing a systematic, gradual decrease of $D(t)$, until a plateau reached at $D_{\infty} \approx 0.053  \pm 0.006$~$\mu$m$^2.s^{-1}$ after approximately $2500$ s. $D_{\infty}$ is independent of the instantaneous bacterial surface density in the range explored here ($10^9$ to $10^{10}$ particles/m$^{2}$). This corresponds to a surface fraction less than $8\%$ and a typical interparticle distance of at least $10$~$\mu$m, larger than the average cells diameter $d \sim 3$~$\mu$m (see Figure~\ref{Base}(b), inset). 

\par Figure~\ref{Fig6}(a) shows the results obtained with the microfluidic system, which is used to renew the population of bacteria in the surface vicinity without changing the surface on which diffusion occurs. After the introduction to the microfluidic cell, the cyanobacteria are allowed to sediment and diffuse onto the surface without applied flow. The observed trend is similar to the experiments in the closed chamber: the diffusion coefficient decreases with time. After about one hour of surface diffusion, the cells are detached from the surface by establishing a fluid flow for twenty minutes (vertical grey bar in Fig.~\ref{Fig6}(a)). Then, the flow is stopped and the cells are allowed to sediment and diffuse onto the surface again. The diffusion coefficient takes the same value as what found just before the washing flow. Since the population of cells was totally renewed by the washing flow, this observation suggests that the surface has retained a signature of the past diffusion events, which influences the behavior of the incoming new cells.  

Moreover, we find that the time evolution of the diffusion coefficient is highly dependent on the ability of \textit{Synechocystis} to produce released EPS, see Figure~\ref{Fig6}(b). The wild type and the single mutant \textit{$\Delta$sll1581} strains, which produce similar high quantities of released EPS as compared to the two double mutant strains (\textit{$\Delta$sll1581-sll1875} and \textit{$\Delta$sll0923-sll5052}), are characterized by the significant decrease of about $60\%$ of the initial diffusion coefficient. In contrast the \textit{$\Delta$sll1581-sll1875} and \textit{$\Delta$sll0923-sll5052} double mutants, which produce much less released EPS, exhibit a smaller decrease of $10\%$ in their diffusion coefficient. Hence, the released EPS are identified as a major cause for the slowdown of the cell dynamics.
\subsection{\label{Permanent}Normal diffusion resulting from twitching motility}
We now focus on the permanent regime once the plateau has been reached, and analyse the trajectories of the bacterial cells for a representative experiment which will be referred to as experiment ``1". Figure~\ref{Base}(c) (inset) displays the linearity of the mean square displacement (Eq.~\ref{eqMSD}) for the representative experiment with a diffusion coefficient at the plateau $D_{\infty,1}=0.059 \pm 0.001$~$\mu$m$^2.s^{-1}$. Yet, the apparently normal surface diffusion of \textit{Synechocystis} stems from the complex dynamics illustrated by the non-Gaussian probability distribution function (PDF) of displacement shown in Figure~\ref{Traj2c}(b). The central part of the distribution corresponds to a state of low mobility (``tumble") while the tails reflect the higher cell mobility occuring during the ``run" periods~\cite{Berg1972}. A typical trajectory in Figure~\ref{Traj2c}(a) indeed reveals twitching motility with run and tumble motion. The run periods are directional and detected with the procedure explained in Appendix~\ref{app_run}, from which we obtain the run times $\tau_{run}$ and additionally the tumble times $\tau_{tumble}$ and the mean velocity of runs, defined as $V_{m}=l_{run}/\tau_{run}$ where $l_{run}$ is the distance travelled during the run. 

We conduct numerical simulations in order to check for the relevance of the detection procedure, see details in Appendix~\ref{app_simu}. First, we fit the experimental distributions of run and tumble time as in Fig.~\ref{P_totales}, Appendix~\ref{app_run}. Additionally, we assume that runs are \textit{ballistic} excursions of length $l_{run}=\tau_{run} \times V_m$, as suggested by experiments. Then, for each step of the simulation, $\tau_{run}$, $\tau_{tumble}$ and  $V_m$ are chosen randomly along the fits of the experimental distributions. This reproduces the experimental displacement PDF as shown by comparing Figure~\ref{Traj2c}(b) and (c), and provides a diffusion coefficient from simulations $D_{simu} \approx D_{\infty,1}$ (inset of Figure~\ref{Base}(c)). 
\begin{figure}
\includegraphics[scale=0.45]{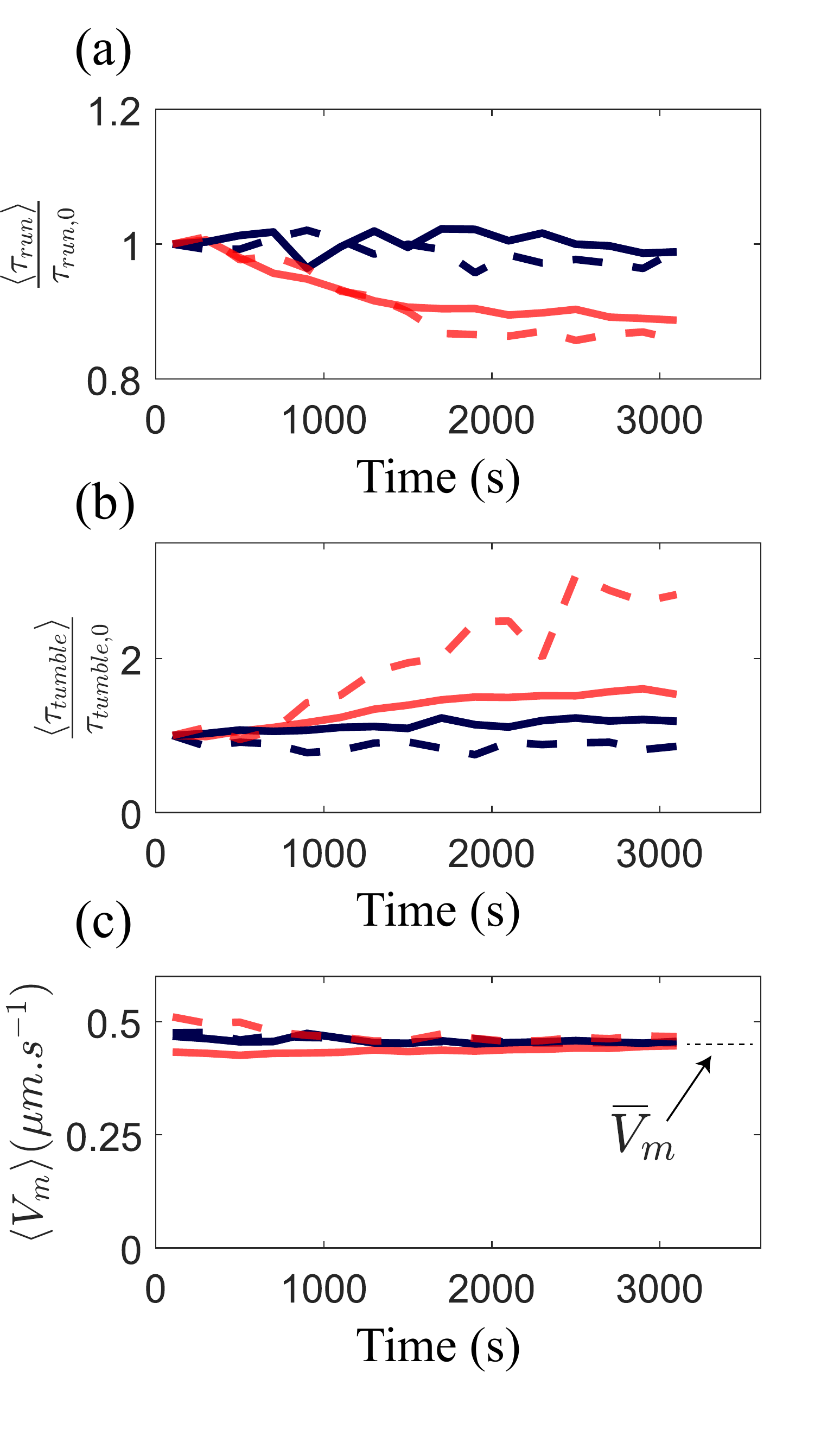}
\caption{\label{labFigrecap_v1} Temporal evolution of (a) $\langle\tau_{run} \rangle$, (b) $\langle \tau_{tumble} \rangle$,  and (c) $\langle V_m \rangle$. $\overline V_m$ is indicated by the arrow. Data are normalized by their initial value for the various \textit{Synechocystis} strains. Same legend as in Figure~\ref{Fig6} (b).}
\end{figure}

Details on the slowdown of motion are provided by analysing the time variation of $\langle \tau_{run} \rangle$, $\langle \tau_{tumble} \rangle$ and  $\langle V_m \rangle$ (here, brackets indicate a time average over temporal windows of $200$~s), plotted Figure~\ref{labFigrecap_v1} (a), (b) and (c) respectively. For the wild type and the $\Delta sll1581$ mutant that produce released EPS, $\langle \tau_{run} \rangle$ decreases and $\langle \tau_{tumble} \rangle$ increases with time. This is not observed with the two double mutants that produce less released EPS and their tumble and run times are almost constant. Importantly, $\langle V_m \rangle$ is not only similar but also constant for all the bacterial strains studied.      
\subsection{\label{Temporal}Surface area covered by trajectories}
\begin{figure*}
\includegraphics[trim = 55mm 0mm 0mm 0mm, clip,scale=0.45]{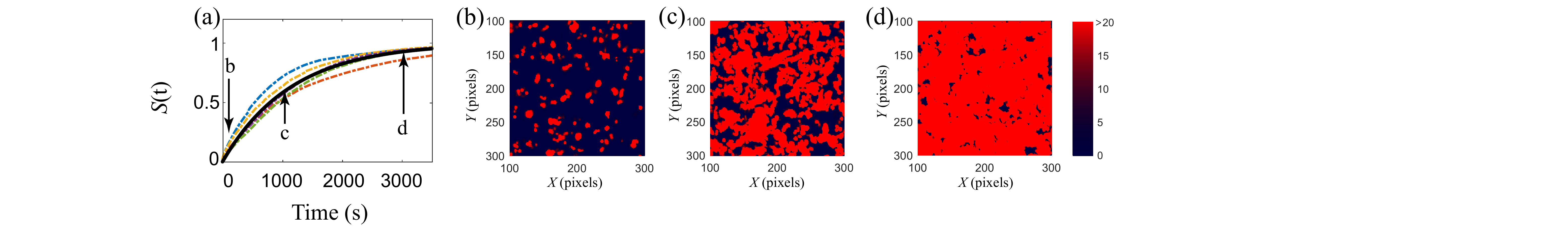}
\caption{\label{Fig4} (a) Fraction of surface area covered by trajectories in various experiments as a function of time (dot-dashed grey lines, colors online). The thick black line is a curve fit with equation~\ref{sdet}, see discussion part. (b), (c) and (d) correspond to colonization maps of the surface for the times indicated in (a). Color maps indicate the cumulated number of visits in the considered pixel (dark: empty site, bright: site visited more than 20 times).}
\end{figure*}
As a basic step for early stages of biofilm formation, we analyse how the surface is explored by the diffusing bacteria. The images are binarised such that the bacteria appear as black disks on a white background, and every black pixel is given a value of $1$ (zero otherwise). The cumulated number of visits for each pixel at a given time is then obtained by summing iteratively all preceding images. Figure~\ref{Fig4} (b-d) displays how the surface area is progressively filled by the trajectories of the diffusing bacteria during the time scale of the experiment. Moreover, although the experiments are conducted in the dilute regime, the fraction of distinct sites (pixels) visited $S(t)$ tends to 1, meaning that the whole surface area of the sample can be screened by EPS. Figure~\ref{Fig4}(a) points out the robustness of this feature.

\section{\label{Discussion}Discussion}
\subsection{Dynamics}
The fact that the average velocity of runs $\langle {V}_m \rangle$ is common to all cyanobacterial strains used in this study implies that it is not linked to the production of EPS, and calls for a simple estimate based on lubrication theory. 

The motion of cyanobacterial cells is ensured by the traction of type IV pili which provides the force $F$ necessary to shear a layer of fluid of viscosity $\eta$ and thickness $h$ squeezed between the bacteria and the surface, leading to $F\sim a \eta V/h$ where $a \sim \pi R^2$ is the sheared area ($R=1.5$~$\mu$m is the radius of a bacterial cell). With $V\sim 1$~$\mu$m.s $^{-1}$ the maximum velocity during runs, $h\sim 1$ nm the thickness of the sheared layer and $\eta \sim 1 \times 10^{-3}$ Pa.s the dynamical viscosity, we obtain the force exerted by the pili $F \sim 7$ pN, consistent with the order of magnitude obtained from mechanical testing with atomic force microscopy \cite{Touhami2006}. The  velocity of the bacterial cells  during run periods is thus limited by the viscous dissipation of the sheared water (liquid mineral medium BG11) layer at the interface between the bacterial cells and the solid surface.
\subsection{Continuous time random walk}
The diffusion coefficient at long times may be obtained by considering a continuous time random walk, where the particle jumps instantaneously over a length $l$ after a waiting time $\tau$~\cite{Montroll1965,Bouchaud1990} as illustrated in the inset of Figure~\ref{Traj2c}(a). We have computed both $l=l_{run}$ and $\tau=  \tau_{run}+\tau_{tumble}$ (taken consecutively) and verified that the second moment of the jump length PDF and the first moment of the waiting time PDF exist. Then, the diffusion coefficient shall take the simple expression $D=\langle l^2 \rangle / 4 \langle \tau \rangle$. Ballistic runs occurring at constant velocity $\overline{V}_{m} \sim 0.47 \pm 0.3$~$\mu$m.s$^{-1}$ (see Fig.~\ref{labFigrecap_v1}(c)) suggest the approximation $\langle l^2 \rangle \approx \overline{V}_{m}^2 \langle \tau_{run}^2 \rangle$. Therefore, an expression for the diffusion coefficient reads 
\begin{equation}
D\sim\frac{1}{4} \overline{V}_m^2 \frac{\langle \tau_{run}^2 \rangle}{\langle \tau \rangle}
\label{eqD}
\end{equation}

The computation of the PDF for $\tau_{\infty,1}$ and $\tau_{run,\infty,1}^2$ leads to well-defined average quantities but without second moment. To calculate $D$ from Eq.~\ref{eqD} we use therefore $\langle \tau_{\infty,1} \rangle  \approx 64$ s and $\langle \tau_{run,\infty,1}^2 \rangle \approx 67$ s$^2$ with both quantities rounded to the nearest whole number and taken at the plateau value for the representative experiment corresponding to $D_{\infty,1}$. This provides $D \approx 0.058$~$\mu$m$^2$.s$^{-1}$ which is similar to experimental values, as indicated in the inset of Figure~\ref{Base}(c). From now on, we shall describe the time evolution of the diffusion coefficient presented in Figure~\ref{Base}(c) with the two parameters $\langle \tau \rangle$ and $\langle \tau_{run}^2 \rangle$.  

The coefficient of diffusion is not affected by inter-particle interactions since experiments are achieved in a dilute regime precluding inter-cell friction and adhesion. Instead, our results with the WT strain and various mutant depleted or not in EPS reveal that the amount of released EPS is key. Moreover, renewal of the population of bacterial cells with the microfluidic setup indicates that the decrease of the diffusion coefficient is due to the modification of the surface properties by previously diffusing cells. In the following, we propose a mechanism whereby released EPS stick on the surface in the form of excreted trails~\cite{Ducret2012,Yu2007}, which modifies the coefficient of diffusion.
\subsection{Mechanism for the slowdown}
\begin{figure}
\includegraphics[scale=0.4]{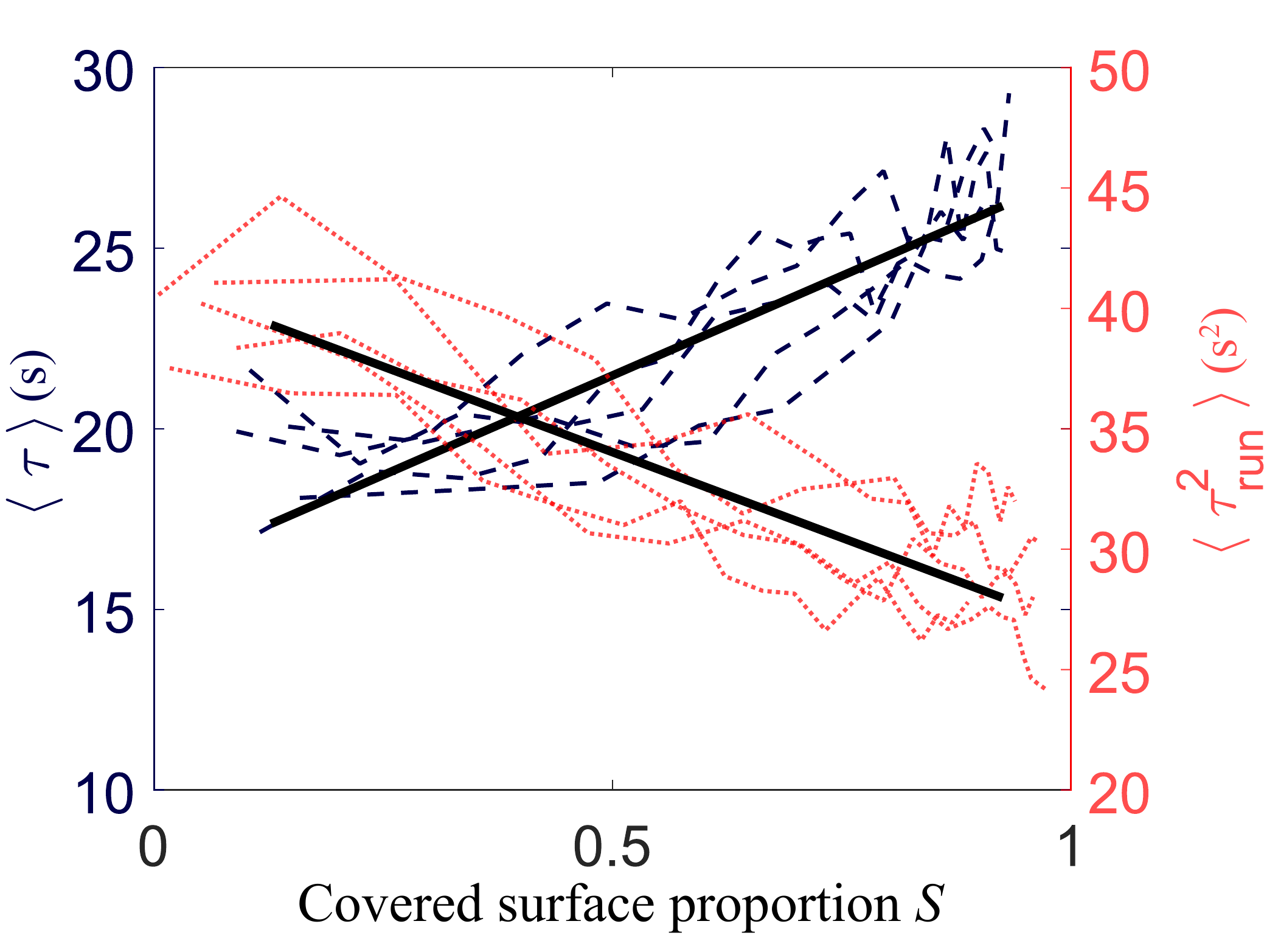}
\caption{\label{Fig5} Variation of $\langle \tau \rangle$ (dashed) and $\langle \tau_{run}^2 \rangle$ (dotted) as functions of the fraction of distinct visited sites $S(t)$, for five different experiments. Plain lines: linear interpolation.}
\end{figure}

The decrease of the coefficient of diffusion and the progressive surface coverage by the trajectories occur on concomitant timescales. Therefore, it is tempting to correlate the surface coverage and the parameters of the model described by Eq.~\ref{eqD}. Figure~\ref{Fig5} shows that $\langle \tau \rangle$ and $\langle\tau_{run}^2\rangle$ vary monotonically with the fraction of distinct sites visited $S(t)$. These two essential temporal parameters, that govern the expression of the diffusion coefficient, may be a function of the probability $P(t)$ to be located on a surface coated by extracellular matrix. By approximating $P(t) \approx S(t)$, both parameters can be written simply as a weighted sum of the covered surface such as $\langle \tau \rangle \approx \langle \tau_{glass} \rangle (1-S(t))+\langle \tau_{eps} \rangle S(t)$, with the same formula for $\tau_{run}^2$ instead of each $\tau$.
\par The computation of $S(t)$ for $N$ random walkers pertains to a class of long standing problems~\cite{Weiss1992}. Here, both the number of diffusing bacterial cells and the coefficient of diffusion are time dependant and the empirical formula
\begin{equation}
\label{sdet}
S(t)=1-e^{-t/\tau_S}
\end{equation}
is used to describe experimental data, resulting in a convenient fit with $\tau_S = 1150 \pm 6$~s $< \tau_N$ (Figure~\ref{Fig4}(a)): most of the surface is visited before sedimentation is complete. 
\par Then, plugging Eq.~\ref{sdet} into the expression for $\langle \tau \rangle$ and $\langle \tau_{run}^2 \rangle$ as above yields a numerical estimate of the temporal evolution of $D(t)$ from Eq.~\ref{eqD}, which can be compared to measurements. Fig.~\ref{Base}(c) indicates a good agreement. Therefore, the present analysis implies that the slowdown of the diffusive dynamics can be reasonably attributed to the coverage of the surface by the excreted exopolysaccharides.
\par Finally, one may ask why cells have longer tumble times and shorter run times on EPS than on glass. Our results show that the decrease of the diffusion coefficient is not related to viscous drag that could result from the shearing of EPS trails left on the surface, since the average velocity during run periods $\overline{V}_m$ remains constant during the experimental timescale. Moreover, $\overline{V}_m$ is similar for the wild type strain and for all mutants regardless of their total EPS production rate, confirming that EPS do not provide additional dissipation during motion. However, it is known that during the early stages of the interaction of the bacterial cells with a surface, cells detect the presence of the extracellular matrix which induces a positive feedback loop that in turn leads to enhancement of EPS production and cell accumulation \cite{Steinberg2015}. For example, \textit{B. subtilis} uses its flagella as mechanosensory organelles for surface sensing. For other micro-organisms such as \textit{Myxococcus xanthus}, EPS play a fundamental role in pilus retraction during social motility \cite{Li2003} and pili mediated twitching motility is affected by surface stiffness, topography and chemistry \cite{Maier2015,Zaburdaev2014,Tuson2013}.

\par Hence, we propose that \textit{Synechocystis} cells sense the EPS deposited on the surface, which triggers cellular changes that affect the temporal characteristics of run and tumble motion. This is consistent with the description of run and tumble rates from linear response theory, for bacteria submitted to spatial changes in concentration of chemicals \cite{DeGennes2004,Segall1986}.

\section{Conclusions}
The experimental results presented here show a correlation between the diffusive dynamics of the bacterial cells and their propensity for released EPS excretion. The observed slowdown of the diffusion is due to the evolution of the characteristic times of the intermittent dynamics, rather than to the enhanced dissipation during the ``runs" due to the viscosity of EPS. Since the estimated surface fraction of visited sites and the coefficient of diffusion evolve with similar timescales, we propose a model for the decrease of the diffusion coefficient based on the deposition of EPS on the surface. This suggests new strategies for controlling biofilm formation, and therefore limiting the infection of host tissues or undesired adhesion in industrial applications.

\begin{acknowledgements}
We thank A. Di Prima and L. Platzer for assistance with the experimental setup and M. Jaharri (LIMSI, Univ. Paris-Sud) and G. Chau for the help with the microfluidic system. We also thank J. Tailleur (MSC, Univ. Paris Diderot) for helpful discussions.
\end{acknowledgements}

\appendix

\section{Calculation of mean square displacement}
\label{app_msd}
\par In order to study the temporal evolution of the motility, we introduce a time-dependent MSD, which is computed at observation time $t$ for different time intervals $\Delta$ as follows: 
\begin{equation}
\label{MSDdt}
\begin{split}
MSD(t,\Delta)= \\
& \hspace{-2.4cm} \frac{1}{N_\delta}\sum_{i=1}^{N_\delta}\frac{1}{2 \delta-\Delta}\sum_{t'=t-\delta+\Delta/2}^{t+\delta-\Delta/2}(X_i(t'+\frac{\Delta}{2})-X_i(t'-\frac{\Delta}{2}))^2
\end{split}
\end{equation}
where $\delta$ is half the time separating two successive values of $t$, and $N_\delta$ is the number of active particles between the times $t-\delta+\Delta/2$ and $t+\delta-\Delta/2$. 

The long time limit of the diffusion is described by computing the ensemble and time-averaged MSD, as defined in Eq.\ref{eqMSD}  where $N$ is the total number of particles, $T_i$ the trajectory length $i$, $X_i(t)$ the position of the particle $i$ at time $t$, and $\Delta$ a given time interval.
\begin{equation}
MSD(\Delta)=\frac{1}{N}\sum_{i=1}^{N}\frac{1}{T_i-\Delta}\sum_{t=1}^{T_i-\Delta}(X_i(t+\Delta)-X_i(t))^2 
\label{eqMSD}
\end{equation}

\section{Run and tumble times}
\label{app_run}
\par Run times $\tau_{run}$ are measured by computing a coarse grained  velocity $V_\delta$ at each time point, with $\delta$ an adjustable  time interval: $$V_\delta(t)=\frac{\left| X(t+\delta/2)-X(t-\delta/2)\right |}{\delta}$$
\begin{figure}
\includegraphics[scale=0.35]{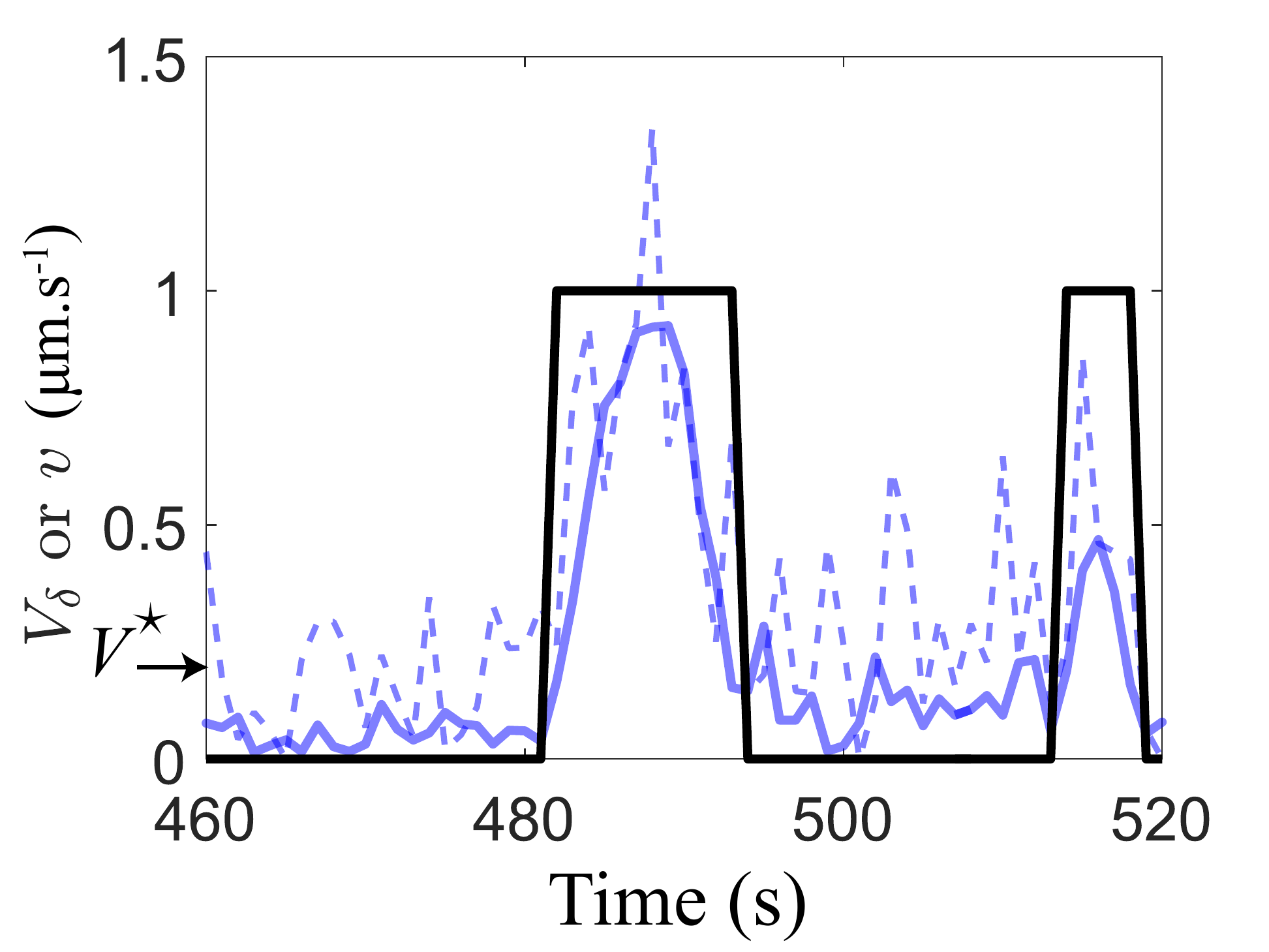}
\caption{\label{Traj2b} Coarse-grained velocity $V_\delta$ (gray plain line) and instantaneous velocity $v$ (gray dashed line). Black plain line indicates the selection of run periods according to the criteria explained in the text (set to 1 when a run is detected, 0 elsewhere). The velocity threshold $V^\star$ is indicated with an arrow.}
\end{figure}
\par An example of such a computation with $\delta=4$ is shown Figure~\ref{Traj2b}, where high velocity spikes, corresponding to runs, are separated by periods of low and noisy signal (tumbles). Run periods are selected according to two different criteria: the velocity $V_\delta>V^\star$ and the distance travelled $\Delta x>\Delta x^\star$ where $V^\star$ and $\Delta x^\star$ are some threshold velocity and length, respectively. Once run periods are accepted, the rest of the trajectory is filled with tumble periods. 

The criteria $V^\star= 0.2$~$\mu$m.s$^{-1}$ and $\Delta x^\star=0.9$~$\mu$m (as suggested by tails of the PDF Figure~\ref{Traj2c}(b)) provide a faithful distinction between the various types of motion. While the choice of $V^\star$ and $\Delta x^\star$ is somewhat arbitrary, the conclusions based on the use of run and tumble times do not depend significantly on these criteria.    
\begin{figure}
\includegraphics[scale=0.3]{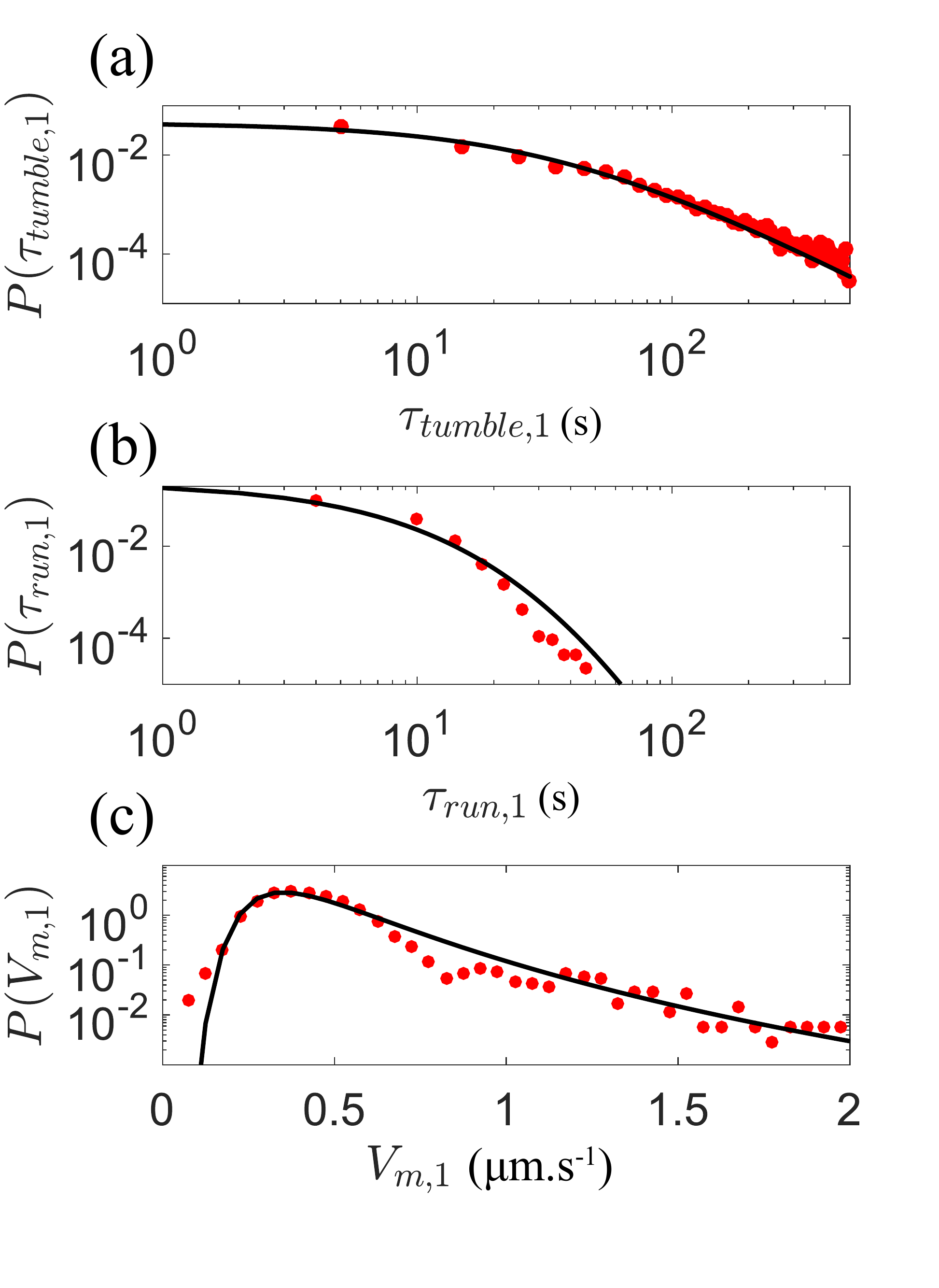}
\caption{\label{P_totales} Distribution of (a) tumble times, (b) run times and (c) average velocity during runs $V_{m}=l_{run}/\tau_{run}$.
Experimental data (points), corresponding to the experiment leading to $D_{\infty,1}$ described in section \ref{Permanent} are fitted with expressions given in Appendix \ref{app_simu} (plain lines), which are used for numerical simulations.}
\end{figure}

\section{Numerical simulations}
\label{app_simu}
Numerical simulations are based on the Monte Carlo Method. Here, $453$ particles are launched, with trajectories of $969$ s. The duration of tumbles and runs is taken from the experimental distributions (Fig.\ref{P_totales}(a) and (b)), fitted by a power law of the form: $$P(\tau_i)=\frac{\alpha_i}{A_i}\frac{1}{(1+\frac{\tau_i}{A_i})^{1+\alpha_i}}$$ with $A_{tumble}=38$~s, $\alpha_{tumble}=1.7$, $A_{run}=42$~s, $\alpha_{run}=11$. The tumble motion is simulated by making a given particle to jiggle in the polar system of coordinates whose center is the fixed position between two runs. The coordinates $(r,\theta)$ are chosen such that $\theta$ is random and $r$ is selected in an exponential distribution of mean $\lambda_{tumble}=0.19$~$\mu$m. The run motion is defined by ballistic excursions of duration $\tau_{run}$, during which the travelled distance is $l_{run}=V_m \tau_{run}$ where $\tau_{run}$ is selected from $P(\tau_{run})$ and $V_m$ is chosen from a Generalized Extreme Value probability law, that conveniently fit the experimental distribution (Fig.~\ref{P_totales}c). This law is defined with a location parameter~$\mu=0.37$~$\mu$m.s$^{-1}$, a scale parameter $\sigma=0.13$~$\mu$m.s$^{-1}$, and a shape parameter $k=0.17$. The angle between two successive runs is random.

%


\begin{thebibliography}{36}%
\makeatletter
\providecommand \@ifxundefined [1]{%
 \@ifx{#1\undefined}
}%
\providecommand \@ifnum [1]{%
 \ifnum #1\expandafter \@firstoftwo
 \else \expandafter \@secondoftwo
 \fi
}%
\providecommand \@ifx [1]{%
 \ifx #1\expandafter \@firstoftwo
 \else \expandafter \@secondoftwo
 \fi
}%
\providecommand \natexlab [1]{#1}%
\providecommand \enquote  [1]{``#1''}%
\providecommand \bibnamefont  [1]{#1}%
\providecommand \bibfnamefont [1]{#1}%
\providecommand \citenamefont [1]{#1}%
\providecommand \href@noop [0]{\@secondoftwo}%
\providecommand \href [0]{\begingroup \@sanitize@url \@href}%
\providecommand \@href[1]{\@@startlink{#1}\@@href}%
\providecommand \@@href[1]{\endgroup#1\@@endlink}%
\providecommand \@sanitize@url [0]{\catcode `\\12\catcode `\$12\catcode
  `\&12\catcode `\#12\catcode `\^12\catcode `\_12\catcode `\%12\relax}%
\providecommand \@@startlink[1]{}%
\providecommand \@@endlink[0]{}%
\providecommand \url  [0]{\begingroup\@sanitize@url \@url }%
\providecommand \@url [1]{\endgroup\@href {#1}{\urlprefix }}%
\providecommand \urlprefix  [0]{URL }%
\providecommand \Eprint [0]{\href }%
\providecommand \doibase [0]{http://dx.doi.org/}%
\providecommand \selectlanguage [0]{\@gobble}%
\providecommand \bibinfo  [0]{\@secondoftwo}%
\providecommand \bibfield  [0]{\@secondoftwo}%
\providecommand \translation [1]{[#1]}%
\providecommand \BibitemOpen [0]{}%
\providecommand \bibitemStop [0]{}%
\providecommand \bibitemNoStop [0]{.\EOS\space}%
\providecommand \EOS [0]{\spacefactor3000\relax}%
\providecommand \BibitemShut  [1]{\csname bibitem#1\endcsname}%
\let\auto@bib@innerbib\@empty
\bibitem [{\citenamefont {Singh}\ \emph {et~al.}(2017)\citenamefont {Singh},
  \citenamefont {Singh}, \citenamefont {Chowdhury},\ and\ \citenamefont
  {Singh}}]{Singh2017}%
  \BibitemOpen
  \bibfield  {author} {\bibinfo {author} {\bibfnamefont {S.}~\bibnamefont
  {Singh}}, \bibinfo {author} {\bibfnamefont {S.~K.}\ \bibnamefont {Singh}},
  \bibinfo {author} {\bibfnamefont {I.}~\bibnamefont {Chowdhury}}, \ and\
  \bibinfo {author} {\bibfnamefont {R.}~\bibnamefont {Singh}},\ }\href
  {\doibase 10.2174/1874285801711010053} {\bibfield  {journal} {\bibinfo
  {journal} {The Open Microbiology Journal}\ }\textbf {\bibinfo {volume}
  {11}},\ \bibinfo {pages} {53} (\bibinfo {year} {2017})}\BibitemShut {NoStop}%
\bibitem [{\citenamefont {Tuson}\ and\ \citenamefont
  {Weibel}(2013)}]{Tuson2013}%
  \BibitemOpen
  \bibfield  {author} {\bibinfo {author} {\bibfnamefont {H.~H.}\ \bibnamefont
  {Tuson}}\ and\ \bibinfo {author} {\bibfnamefont {D.~B.}\ \bibnamefont
  {Weibel}},\ }\href {\doibase 10.1039/c3sm27705d} {\bibfield  {journal}
  {\bibinfo  {journal} {Soft Matter}\ }\textbf {\bibinfo {volume} {9}},\
  \bibinfo {pages} {4368} (\bibinfo {year} {2013})}\BibitemShut {NoStop}%
\bibitem [{\citenamefont {Davies}(2003)}]{Davies2003}%
  \BibitemOpen
  \bibfield  {author} {\bibinfo {author} {\bibfnamefont {D.}~\bibnamefont
  {Davies}},\ }\href {\doibase 10.1038/nrd1008} {\bibfield  {journal} {\bibinfo
   {journal} {Nature Reviews Drug Discovery}\ }\textbf {\bibinfo {volume}
  {2}},\ \bibinfo {pages} {114} (\bibinfo {year} {2003})}\BibitemShut {NoStop}%
\bibitem [{\citenamefont {Costerton}(1999)}]{Costerton1999}%
  \BibitemOpen
  \bibfield  {author} {\bibinfo {author} {\bibfnamefont {J.~W.}\ \bibnamefont
  {Costerton}},\ }\href {\doibase 10.1126/science.284.5418.1318} {\bibfield
  {journal} {\bibinfo  {journal} {Science}\ }\textbf {\bibinfo {volume}
  {284}},\ \bibinfo {pages} {1318} (\bibinfo {year} {1999})}\bibitem [{\citenamefont {Omar}\ \emph {et~al.}(2017)\citenamefont {Omar},
  \citenamefont {Wright}, \citenamefont {Schultz}, \citenamefont {Burrell},\
  and\ \citenamefont {Nadworny}}]{Omar2017}%
  \BibitemOpen
  \bibfield  {author} {\bibinfo {author} {\bibfnamefont {A.}~\bibnamefont
  {Omar}}, \bibinfo {author} {\bibfnamefont {J.}~\bibnamefont {Wright}},
  \bibinfo {author} {\bibfnamefont {G.}~\bibnamefont {Schultz}}, \bibinfo
  {author} {\bibfnamefont {R.}~\bibnamefont {Burrell}}, \ and\ \bibinfo
  {author} {\bibfnamefont {P.}~\bibnamefont {Nadworny}},\ }\href {\doibase
  10.3390/microorganisms5010009} {\bibfield  {journal} {\bibinfo  {journal}
  {Microorganisms}\ }\textbf {\bibinfo {volume} {5}},\ \bibinfo {pages} {9}
  (\bibinfo {year} {2017})}\BibitemShut {NoStop}%
\bibitem [{\citenamefont {Mazza}(2016)}]{Mazza2016}%
  \BibitemOpen
  \bibfield  {author} {\bibinfo {author} {\bibfnamefont {M.~G.}\ \bibnamefont
  {Mazza}},\ }\href {\doibase 10.1088/0022-3727/49/20/203001} {\bibfield
  {journal} {\bibinfo  {journal} {Journal of Physics D: Applied Physics}\
  }\textbf {\bibinfo {volume} {49}},\ \bibinfo {pages} {203001} (\bibinfo
  {year} {2016})}%
\bibitem [{\citenamefont {Taktikos}\ \emph {et~al.}(2015)\citenamefont
  {Taktikos}, \citenamefont {Lin}, \citenamefont {Stark}, \citenamefont
  {Biais},\ and\ \citenamefont {Zaburdaev}}]{Taktikos2015}%
  \BibitemOpen
  \bibfield  {author} {\bibinfo {author} {\bibfnamefont {J.}~\bibnamefont
  {Taktikos}}, \bibinfo {author} {\bibfnamefont {Y.~T.}\ \bibnamefont {Lin}},
  \bibinfo {author} {\bibfnamefont {H.}~\bibnamefont {Stark}}, \bibinfo
  {author} {\bibfnamefont {N.}~\bibnamefont {Biais}}, \ and\ \bibinfo {author}
  {\bibfnamefont {V.}~\bibnamefont {Zaburdaev}},\ }\href {\doibase
  10.1371/journal.pone.0137661} {\bibfield  {journal} {\bibinfo  {journal}
  {PLoS ONE}\ }\textbf {\bibinfo {volume} {10}},\ \bibinfo {pages} {1}
  (\bibinfo {year} {2015})}\BibitemShut {NoStop}%
\bibitem [{\citenamefont {Vlamakis}\ \emph {et~al.}(2013)\citenamefont
  {Vlamakis}, \citenamefont {Chai}, \citenamefont {Beauregard}, \citenamefont
  {Losick},\ and\ \citenamefont {Kolter}}]{Vlamakis2013}%
  \BibitemOpen
  \bibfield  {author} {\bibinfo {author} {\bibfnamefont {H.}~\bibnamefont
  {Vlamakis}}, \bibinfo {author} {\bibfnamefont {Y.}~\bibnamefont {Chai}},
  \bibinfo {author} {\bibfnamefont {P.}~\bibnamefont {Beauregard}}, \bibinfo
  {author} {\bibfnamefont {R.}~\bibnamefont {Losick}}, \ and\ \bibinfo {author}
  {\bibfnamefont {R.}~\bibnamefont {Kolter}},\ }\href {\doibase
  10.1038/nrmicro2960} {\bibfield  {journal} {\bibinfo  {journal} {Nature
  Reviews Microbiology}\ }\textbf {\bibinfo {volume} {11}},\ \bibinfo {pages}
  {157} (\bibinfo {year} {2013})}%
\bibitem [{\citenamefont {Marshall}\ \emph {et~al.}(1971)\citenamefont
  {Marshall}, \citenamefont {Stout},\ and\ \citenamefont
  {Mitchell}}]{Marshall1971-1}%
  \BibitemOpen
  \bibfield  {author} {\bibinfo {author} {\bibfnamefont {K.~C.}\ \bibnamefont
  {Marshall}}, \bibinfo {author} {\bibfnamefont {R.}~\bibnamefont {Stout}}, \
  and\ \bibinfo {author} {\bibfnamefont {R.}~\bibnamefont {Mitchell}},\ }\href
  {http://mic.microbiologyresearch.org/content/journal/micro/10.1099/00221287-68-3-337}
  {\bibfield  {journal} {\bibinfo  {journal} {Microbiology}\ }\textbf {\bibinfo
  {volume} {68}},\ \bibinfo {pages} {337} (\bibinfo {year} {1971})}\BibitemShut
  {NoStop}%
\bibitem [{\citenamefont {Jittawuttipoka}\ \emph {et~al.}(2013)\citenamefont
  {Jittawuttipoka}, \citenamefont {Planchon}, \citenamefont {Spalla},
  \citenamefont {Benzerara}, \citenamefont {Guyot}, \citenamefont
  {Cassier-Chauvat},\ and\ \citenamefont {Chauvat}}]{Jittawuttipoka2013}%
  \BibitemOpen
  \bibfield  {author} {\bibinfo {author} {\bibfnamefont {T.}~\bibnamefont
  {Jittawuttipoka}}, \bibinfo {author} {\bibfnamefont {M.}~\bibnamefont
  {Planchon}}, \bibinfo {author} {\bibfnamefont {O.}~\bibnamefont {Spalla}},
  \bibinfo {author} {\bibfnamefont {K.}~\bibnamefont {Benzerara}}, \bibinfo
  {author} {\bibfnamefont {F.}~\bibnamefont {Guyot}}, \bibinfo {author}
  {\bibfnamefont {C.}~\bibnamefont {Cassier-Chauvat}}, \ and\ \bibinfo {author}
  {\bibfnamefont {F.}~\bibnamefont {Chauvat}},\ }\href {\doibase
  10.1371/journal.pone.0055564} {\bibfield  {journal} {\bibinfo  {journal}
  {PLoS ONE}\ }\textbf {\bibinfo {volume} {8}},\ \bibinfo {pages}
  {e55564} (\bibinfo {year} {2013})}\BibitemShut {NoStop}%
\bibitem [{\citenamefont {Ursell}\ \emph {et~al.}(2013)\citenamefont {Ursell},
  \citenamefont {Chau}, \citenamefont {Wisen}, \citenamefont {Bhaya},\ and\
  \citenamefont {Huang}}]{Bhaya2013}%
  \BibitemOpen
  \bibfield  {author} {\bibinfo {author} {\bibfnamefont {T.}~\bibnamefont
  {Ursell}}, \bibinfo {author} {\bibfnamefont {R.~M.~W.}\ \bibnamefont {Chau}},
  \bibinfo {author} {\bibfnamefont {S.}~\bibnamefont {Wisen}}, \bibinfo
  {author} {\bibfnamefont {D.}~\bibnamefont {Bhaya}}, \ and\ \bibinfo {author}
  {\bibfnamefont {K.~C.}\ \bibnamefont {Huang}},\ }\href {\doibase
  10.1371/journal.pcbi.1003205} {\bibfield  {journal} {\bibinfo  {journal}
  {PLoS Comput Biol}\ }\textbf {\bibinfo {volume} {9}},\ \bibinfo {pages}
  {e1003205} (\bibinfo {year} {2013})}\BibitemShut {NoStop}%
\bibitem [{\citenamefont {Zhao}\ \emph {et~al.}(2013)\citenamefont {Zhao},
  \citenamefont {Tseng}, \citenamefont {Beckerman}, \citenamefont {Jin},
  \citenamefont {Gibiansky}, \citenamefont {Harrison}, \citenamefont {Luijten},
  \citenamefont {Parsek},\ and\ \citenamefont {Wong}}]{Zhao2013}%
  \BibitemOpen
  \bibfield  {author} {\bibinfo {author} {\bibfnamefont {K.}~\bibnamefont
  {Zhao}}, \bibinfo {author} {\bibfnamefont {B.~S.}\ \bibnamefont {Tseng}},
  \bibinfo {author} {\bibfnamefont {B.}~\bibnamefont {Beckerman}}, \bibinfo
  {author} {\bibfnamefont {F.}~\bibnamefont {Jin}}, \bibinfo {author}
  {\bibfnamefont {M.~L.}\ \bibnamefont {Gibiansky}}, \bibinfo {author}
  {\bibfnamefont {J.~J.}\ \bibnamefont {Harrison}}, \bibinfo {author}
  {\bibfnamefont {E.}~\bibnamefont {Luijten}}, \bibinfo {author} {\bibfnamefont
  {M.~R.}\ \bibnamefont {Parsek}}, \ and\ \bibinfo {author} {\bibfnamefont
  {G.~C.~L.}\ \bibnamefont {Wong}},\ }\href {\doibase 10.1038/nature12155}
  {\bibfield  {journal} {\bibinfo  {journal} {Nature}\ }\textbf {\bibinfo
  {volume} {497}},\ \bibinfo {pages} {388} (\bibinfo {year} {2013})}
\bibitem [{\citenamefont {Gelimson}\ \emph {et~al.}(2016)\citenamefont
  {Gelimson}, \citenamefont {Zhao}, \citenamefont {Lee}, \citenamefont {Kranz},
  \citenamefont {Wong},\ and\ \citenamefont {Golestanian}}]{Gelimson2016}%
  \BibitemOpen
  \bibfield  {author} {\bibinfo {author} {\bibfnamefont {A.}~\bibnamefont
  {Gelimson}}, \bibinfo {author} {\bibfnamefont {K.}~\bibnamefont {Zhao}},
  \bibinfo {author} {\bibfnamefont {C.~K.}\ \bibnamefont {Lee}}, \bibinfo
  {author} {\bibfnamefont {W.~T.}\ \bibnamefont {Kranz}}, \bibinfo {author}
  {\bibfnamefont {G.~C.~L.}\ \bibnamefont {Wong}}, \ and\ \bibinfo {author}
  {\bibfnamefont {R.}~\bibnamefont {Golestanian}},\ }\href {\doibase
  10.1103/PhysRevLett.117.178102} {\bibfield  {journal} {\bibinfo  {journal}
  {Physical Review Letters}\ }\textbf {\bibinfo {volume} {117}},\ \bibinfo
  {pages} {178102} (\bibinfo {year} {2016})}
\bibitem [{\citenamefont {Hu}\ \emph {et~al.}(2016)\citenamefont {Hu},
  \citenamefont {Gibiansky}, \citenamefont {Wang}, \citenamefont {Wang},
  \citenamefont {Lux}, \citenamefont {Li}, \citenamefont {Wong},\ and\
  \citenamefont {Shi}}]{Hu2016}%
  \BibitemOpen
  \bibfield  {author} {\bibinfo {author} {\bibfnamefont {W.}~\bibnamefont
  {Hu}}, \bibinfo {author} {\bibfnamefont {M.~L.}\ \bibnamefont {Gibiansky}},
  \bibinfo {author} {\bibfnamefont {J.}~\bibnamefont {Wang}}, \bibinfo {author}
  {\bibfnamefont {C.}~\bibnamefont {Wang}}, \bibinfo {author} {\bibfnamefont
  {R.}~\bibnamefont {Lux}}, \bibinfo {author} {\bibfnamefont {Y.}~\bibnamefont
  {Li}}, \bibinfo {author} {\bibfnamefont {G.~C.~L.}\ \bibnamefont {Wong}}, \
  and\ \bibinfo {author} {\bibfnamefont {W.}~\bibnamefont {Shi}},\ }\href
  {\doibase 10.1038/srep17790} {\bibfield  {journal} {\bibinfo  {journal}
  {Scientific Reports}\ }\textbf {\bibinfo {volume} {6}},\ \bibinfo {pages}
  {17790} (\bibinfo {year} {2016})}\BibitemShut {NoStop}%
\bibitem [{\citenamefont {B\'enichou}\ \emph {et~al.}(2011)\citenamefont
  {B\'enichou}, \citenamefont {Loverdo}, \citenamefont {Moreau},\ and\
  \citenamefont {Voituriez}}]{Benichou2011}%
  \BibitemOpen
  \bibfield  {author} {\bibinfo {author} {\bibfnamefont {O.}~\bibnamefont
  {B\'enichou}}, \bibinfo {author} {\bibfnamefont {C.}~\bibnamefont {Loverdo}},
  \bibinfo {author} {\bibfnamefont {M.}~\bibnamefont {Moreau}}, \ and\ \bibinfo
  {author} {\bibfnamefont {R.}~\bibnamefont {Voituriez}},\ }\href {\doibase
  10.1103/RevModPhys.83.81} {\bibfield  {journal} {\bibinfo  {journal} {Rev.
  Mod. Phys.}\ }\textbf {\bibinfo {volume} {83}},\ \bibinfo {pages} {81}
  (\bibinfo {year} {2011})}\BibitemShut {NoStop}%
\bibitem [{\citenamefont {Panoff}\ \emph {et~al.}(1988)\citenamefont {Panoff},
  \citenamefont {Priem}, \citenamefont {Morvan},\ and\ \citenamefont
  {Joset}}]{Panoff1988}%
  \BibitemOpen
  \bibfield  {author} {\bibinfo {author} {\bibfnamefont {J.~M.}\ \bibnamefont
  {Panoff}}, \bibinfo {author} {\bibfnamefont {B.}~\bibnamefont {Priem}},
  \bibinfo {author} {\bibfnamefont {H.}~\bibnamefont {Morvan}}, \ and\ \bibinfo
  {author} {\bibfnamefont {F.}~\bibnamefont {Joset}},\ }\href {\doibase
  10.1007/BF00408249} {\bibfield  {journal} {\bibinfo  {journal} {Archives of
  Microbiology}\ }\textbf {\bibinfo {volume} {150}},\ \bibinfo {pages} {558}
  (\bibinfo {year} {1988})}\BibitemShut {NoStop}%
\bibitem [{\citenamefont {Bhaya}\ \emph {et~al.}(2000)\citenamefont {Bhaya},
  \citenamefont {Bianco}, \citenamefont {Bryant},\ and\ \citenamefont
  {Grossman}}]{Bhaya2000}%
  \BibitemOpen
  \bibfield  {author} {\bibinfo {author} {\bibfnamefont {D.}~\bibnamefont
  {Bhaya}}, \bibinfo {author} {\bibfnamefont {N.~R.}\ \bibnamefont {Bianco}},
  \bibinfo {author} {\bibfnamefont {D.}~\bibnamefont {Bryant}}, \ and\ \bibinfo
  {author} {\bibfnamefont {A.}~\bibnamefont {Grossman}},\ }\href {\doibase
  10.1046/j.1365-2958.2000.02068.x} {\bibfield  {journal} {\bibinfo  {journal}
  {Molecular Microbiology}\ }\textbf {\bibinfo {volume} {37}},\ \bibinfo
  {pages} {941} (\bibinfo {year} {2000})}\BibitemShut {NoStop}%
\bibitem [{\citenamefont {Ng}\ \emph {et~al.}(2003)\citenamefont {Ng},
  \citenamefont {Grossman},\ and\ \citenamefont {Bhaya}}]{Bhaya2002}%
  \BibitemOpen
  \bibfield  {author} {\bibinfo {author} {\bibfnamefont {W.-O.}\ \bibnamefont
  {Ng}}, \bibinfo {author} {\bibfnamefont {A.~R.}\ \bibnamefont {Grossman}}, \
  and\ \bibinfo {author} {\bibfnamefont {D.}~\bibnamefont {Bhaya}},\ }\href
  {\doibase 10.1128/JB.185.5.1599-1607.2003} {\bibfield  {journal} {\bibinfo
  {journal} {Journal of Bacteriology}\ }\textbf {\bibinfo {volume} {185}},\
  \bibinfo {pages} {1599} (\bibinfo {year} {2003})}
\bibitem [{\citenamefont {Chau}\ \emph {et~al.}(2015)\citenamefont {Chau},
  \citenamefont {Ursell}, \citenamefont {Wang}, \citenamefont {Huang},\ and\
  \citenamefont {Bhaya}}]{Bhaya2015}%
  \BibitemOpen
  \bibfield  {author} {\bibinfo {author} {\bibfnamefont {R.~M.~W.}\
  \bibnamefont {Chau}}, \bibinfo {author} {\bibfnamefont {T.}~\bibnamefont
  {Ursell}}, \bibinfo {author} {\bibfnamefont {S.}~\bibnamefont {Wang}},
  \bibinfo {author} {\bibfnamefont {K.~C.}\ \bibnamefont {Huang}}, \ and\
  \bibinfo {author} {\bibfnamefont {D.}~\bibnamefont {Bhaya}},\ }\href
  {\doibase 10.1016/j.bpj.2015.01.042} {\bibfield  {journal} {\bibinfo
  {journal} {Biophysical Journal}\ }\textbf {\bibinfo {volume} {108}},\
  \bibinfo {pages} {1623} (\bibinfo {year} {2015})}\BibitemShut {NoStop}%
\bibitem [{\citenamefont {Wilde}\ and\ \citenamefont
  {Mullineaux}(2015)}]{Wilde2015}%
  \BibitemOpen
  \bibfield  {author} {\bibinfo {author} {\bibfnamefont {A.}~\bibnamefont
  {Wilde}}\ and\ \bibinfo {author} {\bibfnamefont {C.~W.}\ \bibnamefont
  {Mullineaux}},\ }\href {\doibase 10.1111/mmi.13242} {\bibfield  {journal}
  {\bibinfo  {journal} {Molecular Microbiology}\ }\textbf {\bibinfo {volume}
  {98}},\ \bibinfo {pages} {998} (\bibinfo {year} {2015})}\BibitemShut
  {NoStop}%
\bibitem [{\citenamefont {Berg}\ \emph {et~al.}()\citenamefont {Berg},
  \citenamefont {Darnton},\ and\ \citenamefont {Jaffe}}]{Lien}%
  \BibitemOpen
  \bibfield  {author} {\bibinfo {author} {\bibfnamefont {H.~C.}\ \bibnamefont
  {Berg}}, \bibinfo {author} {\bibfnamefont {N.}~\bibnamefont {Darnton}}, \
  and\ \bibinfo {author} {\bibfnamefont {J.}~\bibnamefont {Jaffe}},\ }\href
  {http://www.rowland.harvard.edu/labs/bacteria/software/index.php} {\enquote
  {\bibinfo {title} {http://www.rowland.harvard.edu/labs/bacteria/software/index.php}}\ }\BibitemShut {NoStop}%
\bibitem [{\citenamefont {Glaser}\ and\ \citenamefont
  {Higgins}(1989)}]{Glaser1989}%
  \BibitemOpen
  \bibfield  {author} {\bibinfo {author} {\bibfnamefont {D.}~\bibnamefont
  {Glaser}}\ and\ \bibinfo {author} {\bibfnamefont {M.}~\bibnamefont
  {Higgins}},\ }\href {\doibase 10.1128/jb.171.2.669-673.1989} {\bibfield
  {journal} {\bibinfo  {journal} {Journal of Bacteriology}\ }\textbf {\bibinfo
  {volume} {171}},\ \bibinfo {pages} {669} (\bibinfo {year}
  {1989})}\BibitemShut {NoStop}%
\bibitem [{\citenamefont {Berg}\ and\ \citenamefont {Brown}(1972)}]{Berg1972}%
  \BibitemOpen
  \bibfield  {author} {\bibinfo {author} {\bibfnamefont {H.~C.}\ \bibnamefont
  {Berg}}\ and\ \bibinfo {author} {\bibfnamefont {D.~a.}\ \bibnamefont
  {Brown}},\ }\href {\doibase 10.1038/239500a0} {\bibfield  {journal} {\bibinfo
   {journal} {Nature}\ }\textbf {\bibinfo {volume} {239}},\ \bibinfo {pages}
  {500} (\bibinfo {year} {1972})}\BibitemShut {NoStop}%
\bibitem [{\citenamefont {Montroll}\ and\ \citenamefont
  {Weiss}(1965)}]{Montroll1965}%
  \BibitemOpen
  \bibfield  {author} {\bibinfo {author} {\bibfnamefont {E.}~\bibnamefont
  {Montroll}}\ and\ \bibinfo {author} {\bibfnamefont {G.}~\bibnamefont
  {Weiss}},\ }\href {\doibase 10.1063/1.1704269} {\bibfield  {journal}
  {\bibinfo  {journal} {Journal of Mathematical Physics}\ }\textbf {\bibinfo
  {volume} {6}},\ \bibinfo {pages} {167} (\bibinfo {year} {1965})}\BibitemShut
  {NoStop}%
\bibitem [{\citenamefont {Bouchaud}\ and\ \citenamefont
  {Georges}(1990)}]{Bouchaud1990}%
  \BibitemOpen
  \bibfield  {author} {\bibinfo {author} {\bibfnamefont {J.~P.}\ \bibnamefont
  {Bouchaud}}\ and\ \bibinfo {author} {\bibfnamefont {A.}~\bibnamefont
  {Georges}},\ }\href {\doibase 10.1016/0370-1573(90)90099-N} {\emph {\bibinfo
  {title} {Physics Reports}}},\ Vol.\ \bibinfo {volume} {195}\ (\bibinfo {year}
  {1990})\ pp.\ \bibinfo {pages} {127--293}\BibitemShut {NoStop}%
\bibitem [{\citenamefont {Touhami}\ \emph {et~al.}(2006)\citenamefont
  {Touhami}, \citenamefont {Jericho}, \citenamefont {Boyd},\ and\ \citenamefont
  {Beveridge}}]{Touhami2006}%
  \BibitemOpen
  \bibfield  {author} {\bibinfo {author} {\bibfnamefont {A.}~\bibnamefont
  {Touhami}}, \bibinfo {author} {\bibfnamefont {M.~H.}\ \bibnamefont
  {Jericho}}, \bibinfo {author} {\bibfnamefont {J.~M.}\ \bibnamefont {Boyd}}, \
  and\ \bibinfo {author} {\bibfnamefont {T.~J.}\ \bibnamefont {Beveridge}},\
  }\href {\doibase 10.1128/JB.188.2.370-377.2006} {\bibfield  {journal}
  {\bibinfo  {journal} {Journal of Bacteriology}\ }\textbf {\bibinfo {volume}
  {188}},\ \bibinfo {pages} {370} (\bibinfo {year} {2006})}\BibitemShut
  {NoStop}%
\bibitem [{\citenamefont {Ducret}\ \emph {et~al.}(2012)\citenamefont {Ducret},
  \citenamefont {Valignat}, \citenamefont {Mouhamar}, \citenamefont {Mignot},\
  and\ \citenamefont {Theodoly}}]{Ducret2012}%
  \BibitemOpen
  \bibfield  {author} {\bibinfo {author} {\bibfnamefont {A.}~\bibnamefont
  {Ducret}}, \bibinfo {author} {\bibfnamefont {M.-P.}\ \bibnamefont
  {Valignat}}, \bibinfo {author} {\bibfnamefont {F.}~\bibnamefont {Mouhamar}},
  \bibinfo {author} {\bibfnamefont {T.}~\bibnamefont {Mignot}}, \ and\ \bibinfo
  {author} {\bibfnamefont {O.}~\bibnamefont {Theodoly}},\ }\href {\doibase
  10.1073/pnas.1120979109} {\bibfield  {journal} {\bibinfo  {journal}
  {Proceedings of the National Academy of Sciences}\ }\textbf {\bibinfo
  {volume} {109}},\ \bibinfo {pages} {10036} (\bibinfo {year}
  {2012})}\BibitemShut {NoStop}%
\bibitem [{\citenamefont {Yu}\ and\ \citenamefont {Kaiser}(2007)}]{Yu2007}%
  \BibitemOpen
  \bibfield  {author} {\bibinfo {author} {\bibfnamefont {R.}~\bibnamefont
  {Yu}}\ and\ \bibinfo {author} {\bibfnamefont {D.}~\bibnamefont {Kaiser}},\
  }\href {\doibase 10.1111/j.1365-2958.2006.05536.x} {\bibfield  {journal}
  {\bibinfo  {journal} {Molecular Microbiology}\ }\textbf {\bibinfo {volume}
  {63}},\ \bibinfo {pages} {454} (\bibinfo {year} {2007})}\BibitemShut
  {NoStop}%
  \bibitem [{\citenamefont {Larralde}\ and\ \citenamefont {Weiss}(1992)}]{Weiss1992}%
  \BibitemOpen
  \bibfield  {author} {\bibinfo {author} {\bibfnamefont {H.}~\bibnamefont
  {Larralde}}\ and\ \bibinfo {author} {\bibfnamefont {P.}~\bibnamefont {Trunfio}}\ and\ \bibinfo {author} {\bibfnamefont {H.E.}~\bibnamefont {Stanley}}\ and\ \bibinfo {author} {\bibfnamefont {G.H.}~\bibnamefont {Weiss}},\
  }\href {\doibase 10.1038/355423a0} {\bibfield  {journal}
  {\bibinfo  {journal} {Nature}\ }\textbf {\bibinfo {volume}
  {335}},\ \bibinfo {pages} {423} (\bibinfo {year} {1992})}\BibitemShut
  {NoStop}%
\bibitem [{\citenamefont {Rossi}\ and\ \citenamefont {{De
  Philippis}}(2015)}]{Rossi2015}%
  \BibitemOpen
  \bibfield  {author} {\bibinfo {author} {\bibfnamefont {F.}~\bibnamefont
  {Rossi}}\ and\ \bibinfo {author} {\bibfnamefont {R.}~\bibnamefont {{De
  Philippis}}},\ }\href {\doibase 10.3390/life5021218} {\bibfield  {journal}
  {\bibinfo  {journal} {Life}\ }\textbf {\bibinfo {volume} {5}},\ \bibinfo
  {pages} {1218} (\bibinfo {year} {2015})}\BibitemShut {NoStop}%
\bibitem [{\citenamefont {Steinberg}\ and\ \citenamefont
  {Kolodkin-Gal}(2015)}]{Steinberg2015}%
  \BibitemOpen
  \bibfield  {author} {\bibinfo {author} {\bibfnamefont {N.}~\bibnamefont
  {Steinberg}}\ and\ \bibinfo {author} {\bibfnamefont {I.}~\bibnamefont
  {Kolodkin-Gal}},\ }\href {\doibase 10.1128/JB.02516-14} {\bibfield  {journal}
  {\bibinfo  {journal} {Journal of Bacteriology}\ }\textbf {\bibinfo {volume}
  {197}},\ \bibinfo {pages} {2092} (\bibinfo {year} {2015})}\BibitemShut
  {NoStop}%
\bibitem [{\citenamefont {Li}\ \emph {et~al.}(2003)\citenamefont {Li},
  \citenamefont {Sun}, \citenamefont {Ma}, \citenamefont {Lu}, \citenamefont
  {Lux}, \citenamefont {Zusman},\ and\ \citenamefont {Shi}}]{Li2003}%
  \BibitemOpen
  \bibfield  {author} {\bibinfo {author} {\bibfnamefont {Y.}~\bibnamefont
  {Li}}, \bibinfo {author} {\bibfnamefont {H.}~\bibnamefont {Sun}}, \bibinfo
  {author} {\bibfnamefont {X.}~\bibnamefont {Ma}}, \bibinfo {author}
  {\bibfnamefont {A.}~\bibnamefont {Lu}}, \bibinfo {author} {\bibfnamefont
  {R.}~\bibnamefont {Lux}}, \bibinfo {author} {\bibfnamefont {D.}~\bibnamefont
  {Zusman}}, \ and\ \bibinfo {author} {\bibfnamefont {W.}~\bibnamefont {Shi}},\
  }\href {\doibase 10.1073/pnas.0836639100} {\bibfield  {journal} {\bibinfo
  {journal} {Proceedings of the National Academy of Sciences of the United
  States of America}\ }\textbf {\bibinfo {volume} {100}},\ \bibinfo {pages}
  {5443} (\bibinfo {year} {2003})}\BibitemShut {NoStop}%
\bibitem [{\citenamefont {Maier}\ and\ \citenamefont {Wong}(2015)}]{Maier2015}%
  \BibitemOpen
  \bibfield  {author} {\bibinfo {author} {\bibfnamefont {B.}~\bibnamefont
  {Maier}}\ and\ \bibinfo {author} {\bibfnamefont {G.~C.~L.}\ \bibnamefont
  {Wong}},\ }\href {\doibase 10.1016/j.tim.2015.09.002} {\bibfield  {journal}
  {\bibinfo  {journal} {Trends in Microbiology}\ }\textbf {\bibinfo {volume}
  {23}},\ \bibinfo {pages} {775} (\bibinfo {year} {2015})}\BibitemShut
  {NoStop}%
\bibitem [{\citenamefont {Zaburdaev}\ \emph {et~al.}(2014)\citenamefont
  {Zaburdaev}, \citenamefont {Biais}, \citenamefont {Schmiedeberg},
  \citenamefont {Eriksson}, \citenamefont {Jonsson}, \citenamefont {Sheetz},\
  and\ \citenamefont {Weitz}}]{Zaburdaev2014}%
  \BibitemOpen
  \bibfield  {author} {\bibinfo {author} {\bibfnamefont {V.}~\bibnamefont
  {Zaburdaev}}, \bibinfo {author} {\bibfnamefont {N.}~\bibnamefont {Biais}},
  \bibinfo {author} {\bibfnamefont {M.}~\bibnamefont {Schmiedeberg}}, \bibinfo
  {author} {\bibfnamefont {J.}~\bibnamefont {Eriksson}}, \bibinfo {author}
  {\bibfnamefont {A.~B.}\ \bibnamefont {Jonsson}}, \bibinfo {author}
  {\bibfnamefont {M.~P.}\ \bibnamefont {Sheetz}}, \ and\ \bibinfo {author}
  {\bibfnamefont {D.~A.}\ \bibnamefont {Weitz}},\ }\href {\doibase
  10.1016/j.bpj.2014.07.061} {\bibfield  {journal} {\bibinfo  {journal}
  {Biophysical Journal}\ }\textbf {\bibinfo {volume} {107}},\ \bibinfo {pages}
  {1523} (\bibinfo {year} {2014})}\BibitemShut {NoStop}%
\bibitem [{\citenamefont {{De Gennes}}(2004)}]{DeGennes2004}%
  \BibitemOpen
  \bibfield  {author} {\bibinfo {author} {\bibfnamefont {P.-G.}\ \bibnamefont
  {{De Gennes}}},\ }\href {\doibase 10.1007/s00249-004-0426-z} {\bibfield
  {journal} {\bibinfo  {journal} {European Biophysics Journal}\ }\textbf
  {\bibinfo {volume} {33}},\ \bibinfo {pages} {691} (\bibinfo {year}
  {2004})}\BibitemShut {NoStop}%
\bibitem [{\citenamefont {Segall}\ \emph {et~al.}(1986)\citenamefont {Segall},
  \citenamefont {Block},\ and\ \citenamefont {Berg}}]{Segall1986}%
  \BibitemOpen
  \bibfield  {author} {\bibinfo {author} {\bibfnamefont {J.~E.}\ \bibnamefont
  {Segall}}, \bibinfo {author} {\bibfnamefont {S.~M.}\ \bibnamefont {Block}}, \
  and\ \bibinfo {author} {\bibfnamefont {H.~C.}\ \bibnamefont {Berg}},\ }\href
  {https://www.ncbi.nlm.nih.gov/pubmed/3024160} {\ \textbf {\bibinfo {volume}
  {83}},\ \bibinfo {pages} {8987} (\bibinfo {year} {1986})}\BibitemShut
  {NoStop}%
\end{thebibliography}
\end{document}